# Using Targeted Maximum Likelihood Estimation to Estimate Treatment Effect with Longitudinal Continuous or Binary Data: A Systematic Evaluation of 28 Diabetes Clinical Trials

**Lingjing Jiang[1], Michael Rosenblum[2], Yu Du[3]**

[1]Janssen Research & Development, Spring House, PA 19477, USA

[2]Department of Biostatistics, John Hopkins Bloomberg School of Public Health, Baltimore, MD 21205, USA

[3]Statistics, Data and Analytics, Eli Lilly and Company, Indianapolis, IN 46225, USA

*email:* du_yu@lilly.com

SUMMARY:   The primary analysis of clinical trials in diabetes therapeutic area often involves a mixed-model repeated measure (MMRM) approach to estimate the average treatment effect for longitudinal continuous outcome, and a generalized linear mixed model (GLMM) approach for longitudinal binary outcome. In this paper, we considered another estimator of the average treatment effect, called targeted maximum likelihood estimator (TMLE). This estimator can be a one-step alternative to model either continuous or binary outcome. We compared those estimators by simulation studies and by analyzing real data from 28 diabetes clinical trials. The simulations involved different missing data scenarios, and the real data sets covered a wide range of possible distributions of the outcome and covariates in real-life clinical trials for diabetes drugs with different mechanisms of action. For all the settings, adjusted estimators tended to be more efficient than the unadjusted one. In the setting of longitudinal continuous outcome, the MMRM approach with visits and baseline variables interaction appeared to dominate the performance of the MMRM considering the main effects only for the baseline variables while showing better or comparable efficiency to the TMLE estimator in both simulations and data applications. For modeling longitudinal binary outcome, TMLE generally outperformed GLMM in terms of relative efficiency, and its avoidance of the cumbersome covariance fitting procedure from GLMM makes TMLE a more advantageous estimator.

KEY WORDS:   targeted maximum likelihood estimation, MMRM, diabetes, randomized trial

This paper has been submitted for consideration for publication in *Biometrics*

## 1. Introduction

In randomized clinical trials from disease areas like diabetes, Alzheimer, autoimmune, etc., the mixed-model repeated measure (MMRM) approach is often used in the primary analysis for longitudinal continuous outcome, and the generalized linear mixed model (GLMM) is commonly used for longitudinal binary outcome (Chen et al., 2018; Forman et al., 2018; O'Neil et al., 2018; Grunberger et al., 2020). These methods for estimating the average treatment effect adjust for potential bias due to patient dropout by leverage information in baseline variables and post-baseline longitudinal measurements of the outcome. In this paper we considered another statistical method to estimate the average treatment effect, called targeted maximum likelihood estimator (TMLE) that has been extended to handle the setting with longitudinal outcome (Van der Laan and Gruber, 2012; Rosenblum et al., 2018). The TMLE estimator has been shown to have theoretical merits over MMRM or GLMM, due to its doubly robust property. Rosenblum et al. (2018) has conducted head-to-head comparisons between TMLE versus other commonly used estimation methods (e.g. ANCOVA, PLEASE, MMRM) in estimating the average treatment effect with longitudinal continuous outcome from a simulation study mimicking a Alzheimer clinical trial for the treatment against mild cognitive impairment, which showed that the TMLE estimator appeared to have greater robustness to model misspecification i.e., less bias, and in some settings higher precision as well compared to other estimators. We expanded this work by conducting a systematic evaluation and comparison in both simulation and real data applications with 28 pivotal clinical trials from diabetes therapeutic area. We also extended the outcome of interest from longitudinal continuous outcome to including longitudinal binary outcome. In addition to the TMLE estimator, we evaluated two commonly used MMRM approaches in the analyses of diabetes clinical trials. Those are MMRM approach with visits and baseline variables interaction (denoted by MMRM*) and the MMRM approach considering the main effects only for the baseline variables (denoted by MMRM in the following text). The systematic review in the paper provides compelling data evidence and is expected to offer insight to the recommendation of estimation approach for the average treatment effect from diabetes clinical trials with longitudinal continuous or binary outcome.

In this paper, we compared the performance of MMRM, MMRM*, GLMM and TMLE in both simulations and 28 diabetes clinical trials in estimating the average treatment effect on the reduction in HbA1c from baseline for continuous outcome and on the proportion of participants reaching clinical HbA1c targets for binary outcomes at the end of the treatment period. The data on HbA1c was collected longitudinally throughout the trial period. In Section 2, we describe the 28 diabetes studies, which cover a wide range of possible distributions of outcome and covariates that we are expected to observe in real life clinical trials. In Section 3, we define the estimators of the interest on the average treatment effect. In Section 4, we conduct simulation studies that mimic key features from a diabetes pivotal study. In Section 5, we present our main results from clinical trial applications. In section 6, we offer some practical recommendations for estimating treatment effect with longitudinal continuous or binary data.

## 2. Twenty-Eight Diabetes Randomized Clinical Trials

In this paper, we evaluated and compared TMLE vs MMRM/GLMM in 28 phase 3 randomized clinical trials investigating the effects of the treatments on the reduction in HbA1c from baseline and on the proportion of participants reaching clinical HbA1c targets at the end of the treatment period. These phase 3 clinical trials covered multiple classes of diabetes drugs including insulin and non-insulin compounds. There are two outcomes of interests when it comes to testing drug's glycemic control:

(1) Change from baseline in HbA1c at the end of the study treatment period

HbA1c provides an estimate of blood glucose values averaged over a three-month period. It is continuous-valued, so is change in HbA1c from baseline. These values are collected longitudinally at scheduled visits throughout the study duration.

(2) Whether a participant reaches a HbA1c target at the end of the study treatment period

Another outcome of interest related to HbA1c is whether a participant reaches a HbA1c target, namely whether the HbA1c falls below a certain threshold at the end of the study treatment period. For example, HbA1c $< 7\%$ (Davies et al., 2018) and HbA1c $\leqslant 6.5\%$ (Garber et al., 2013) are two commonly used HbA1c targets. This outcome of interest is binary-valued, which is transformed from the continuous-valued HbA1c with the selected target (the 7% threshold is used throughout the clinical trials in this paper). Thus it is also longitudinal in nature.

The sample sizes of these studies ranged from 182 to 1517 (with mean at 563 and SD of 363) while the missing percentage for the primary outcome at the last visit ranged from 4% to 52% (with mean at 15% and SD of 10%). We evaluated the performance of TMLE in estimating the treatment contrasts defined by each of the outcomes of interest and compared with the standard approaches. For continuous-valued longitudinal data, MMRM/MMRM* is often applied to estimate the treatment effect while for binary-valued longitudinal data, a GLMM approach is preferred. The details of these methods are provided in the Section 3.

## 3. Estimator Definitions

### 3.1 *Estimators of Average Treatment Effect*

We focus on the randomized clinical trials where participants have measurements recorded at baseline and $K$ post-baseline visits. For example, as mentioned earlier, in diabetes clinical trials, the primary endpoint is usually the change from baseline in HbA1c which is collected longitudinally. Each participant contributes the generic data vector $(\boldsymbol{X}, A, \boldsymbol{Y}, \boldsymbol{R})$, where $\boldsymbol{X}$ is a $M$ by 1 vector of predefined baseline variables (including baseline outcome of interest), $A$ is the study arm assignment, $\boldsymbol{Y}$ is a $K$ by 1 vector of continuous or binary post-baseline outcome, and $\boldsymbol{R}$ is a $K$ by 1 vector that records the missingness of post-baseline outcomes. We assume the study arm assignment indicator $A$ is binary ($A = 1$ for treatment and $A = 0$ for control). The components of the baseline vector $\boldsymbol{X}$ can be continuous, binary, ordinal, and/or categorical. We assume that the components of $(1, \boldsymbol{X}')$ are linearly independent, that is, no component is a linear combination of the others. We also assume that the baseline vector $\boldsymbol{X}$ is independent of the study assignment $A$, which holds by design since study arm assignment is randomized.

Each participant may be censored due to loss to follow-up (dropout). Let $R_t$ be the indicator variable that the participant attends the next visit, i.e., the visit at time $t + 1$, then $R_t = 1$ if $Y_{t+1}$ is observed, and $R_t = 0$ if otherwise. We assume monotone censoring (i.e. right censoring) in the dropout, that is, if $R_t = 0$, then $R_{t+1} = 0$. We make the missing at random assumption, that is participant dropout is independent of the primary outcome conditioned on the observed data at all visits before dropout.

The goal is to estimate the average treatment effect $\Delta = E[Y_K | A = 1] - E[Y_K | A = 0]$, that is, the difference between populations means at the last visit of the treatment period if everyone in the study population had been assigned to treatment versus control.

Denote the unadjusted estimator (which ignores baseline variables) of the average treatment effect $\Delta$ at final visit $K$ as

$$\hat{\Delta}_{unadj} = \frac{\sum_{i=1}^{N} Y_{i,K} R_{i,K-1} A_i}{\sum_{i=1}^{N} R_{i,K-1} A_i} - \frac{\sum_{i=1}^{N} Y_{i,K} R_{i,K-1}(1 - A_i)}{\sum_{i=1}^{N} R_{i,K-1}(1 - A_i)} \tag{1}$$

### 3.2 *MMRM Estimator - Continuous Outcome*

MMRM estimator is commonly used to estimate the average treatment effect for continuous outcome in the analyses of diabetes clinical trials, which specifies a longitudinal linear model for the vector of outcomes $Y_{i,1}, ..., Y_{i,K}$ for participants $i = 1, ..., N$ at post-baseline visit

$t = 1, ..., K$ where

$$Y_{i,t} = \beta_1 + \sum_{j=2}^{K} \beta_j I(t = j) + \alpha_1 I(A_i = 1) + \sum_{j=2}^{K} \alpha_j I(t = j) \times I(A_i = 1) + \boldsymbol{\beta_X}' \boldsymbol{X_i} + \epsilon_{i,t}, \quad (2)$$

and $\epsilon_{i,t} \sim N(0, \sigma_t^2)$ with $Corr(\epsilon_{i,t}, \epsilon_{i,t'}) = \rho_{t,t'}$ for $t \neq t'$. The indicator function, $I(x)$, is defined as 1 if $x$ is true and 0 otherwise. $\boldsymbol{X}$ is a vector of baseline variables, and $\boldsymbol{\beta_X}$ is the corresponding vector of regression coefficients.

The MMRM estimator of the average treatment effect at final visit $K$ is

$$\hat{\Delta}_{MMRM} = \alpha_1 + \alpha_K. \quad (3)$$

### 3.3  *MMRM\* Estimator - Continuous Outcome*

MMRM\* estimator differs from MMRM estimator in that it includes the interaction between visits and baseline variables besides the main effects from the baseline variables, therefore,

$$Y_{i,t} = \theta_1 + \sum_{j=2}^{K} \theta_j I(t = j) + \gamma_1 I(A_i = 1) + \sum_{j=2}^{K} \gamma_j I(t = j) \times I(A_i = 1) \quad (4)$$
$$+ \sum_{j=2}^{K} \boldsymbol{\beta_{Xj}}' I(t = j) \boldsymbol{X_i} + \epsilon_{i,t},$$

The MMRM\* estimator of the average treatment effect at final visit $K$ is

$$\hat{\Delta}_{MMRM^*} = \gamma_1 + \gamma_K. \quad (5)$$

Due to the added interaction terms, MMRM\* is able to capture the timy-varying correlation between the baseline variables and the outcomes across visits. This approach has also been recommended for the primary analysis by Mallinckrodt et al. (2008, 2020). The estimators defined in (3) and (5) are consistent regardless of model misspecification when dropout is completely at random or when the longitudinal linear model is correctly specified if dropout is missing at random.

### 3.4  *Generalized Linear Mixed Models - Binary Outcome*

Generalized linear mixed models (GLMM), implemented as Proc GLIMMIX in SAS, is a widely-used method for estimating treatment effect in randomized clinical trials at the last visit with longitudinal binary outcome (Liang and Zeger, 1986; Zeger and Liang, 1986). The GLMM approach specifies a marginal longitudinal logistic regression model as

$$logit(P(Y_{i,K} = 1 | A_i = 1, \boldsymbol{X_i})) = \beta_1 + \sum_{j=2}^{K} \beta_j I(t=j) + \alpha_1 I(A_i = 1) \tag{6}$$
$$+ \sum_{j=2}^{K} \alpha_j I(t=j) \times I(A_i = 1) + \boldsymbol{\beta_X}' \boldsymbol{X_i},$$

with a covariance matrix most likely specified in an unstructured format. To estimate the mean response for each treatment group at the final visit, a conditional approach is widely used in practice, where $\bar{\boldsymbol{X}}$, the estimated population mean for the baseline covariates assuming continuous-valued replaced $\boldsymbol{X_i}$ such that the estimated mean response is $E[\widehat{Y_{i,t} | A_i}, \bar{\boldsymbol{X}}]$ and its difference between the treatment groups estimates the average treatment effect. This interpretation of treatment effect is problematic because it is the mean response difference estimated at the mean value of the baseline covariates $\bar{\boldsymbol{X}}$ for the studied population, which is generally not equal to the treatment difference for the overall targeted population. This results in the estimation of a conditional average treatment effect that requires the model (6) to be correctly specified, which can hardly be the case in practice.

Most of the time, the sponsor is interested in estimating the unconditional treatment effect with binary outcome. Therefore, we considered the unconditional average treatment effect defined by the risk difference comparing treated group to untreated group. After fitting the model in (6), we estimated the predicted response for each participant under both treatment group and control group respectively. The average of those estimates across the entire set of participants under each treatment group is defined as the mean response for the group. The difference between these two averages provided the estimated average treatment effect. This estimator is given in (7). Such construct of this estimator has been recommended by FDA in estimating the unconditional average treatment effect with binary outcomes in a draft guidance on covariate adjustment  (FDA Draft Guidance, 2021).

$$\hat{\Delta}_{GLMM} = \frac{1}{N}\sum_{i=1}^{N}P(Y_{i,K}=1|A_i=0,\boldsymbol{X_i}) \tag{7}$$

$$-\frac{1}{N}\sum_{i=1}^{N}P(Y_{i,K}=1|A_i=1,\boldsymbol{X_i})$$

$$=\frac{1}{N}\sum_{i=1}^{N}logit^{-1}(\beta_1+\beta_K+\boldsymbol{\beta_X'X_i})$$

$$-\frac{1}{N}\sum_{i=1}^{N}logit^{-1}(\beta_1+\beta_K+\alpha_1+\alpha_K+\boldsymbol{\beta_X'X_i})$$

The covariance matrix is often modeled through an unstructured form, like those in MMRM. However, if model with unstructured covariance matrix does not converge, analysts will move to a simpler structure, like autoregression with order 1 (AR1), compound symmetry, or some other structures. This process will go on until convergence is achieved.

### 3.5  *Targeted Maximum Likelihood Estimator*

Targeted Maximum Likelihood Estimator (TMLE) can be a one-stop alternative to model either continuous or binary outcome. TMLE is a doubly robust maximum-likelihood-based method that optimizes the bias-variance tradeoff for the target parameter at the expense of the estimation of nuisance parameters (Van Der Laan and Rubin, 2006). The TMLE estimator for longitudinal data is a special case of the sequential regression targeted minimum loss based estimator (Van der Laan and Gruber, 2012; Rosenblum et al., 2018), which is composed of three main models in the following sequence:

(1) propensity score model $P(A=1|\boldsymbol{X})$

(2) dropout model $P(R_t=1|A,\boldsymbol{X},\underline{Y_t},R_{t-1}=1)$ for $t=0,...,K-1$

(3) outcome models $E[Y_t|A,\boldsymbol{X},Y_{\underline{t-1}}]$ for $t=K,K-1,...,2,$

where $\underline{Y_t}$ represents the history of longitudinally collected outcome variables up to time $t$. $Y_t$ can either be continuous or binary-valued measurement of the outcome at time $t$. These are called working models since we do not assume the true, unknown data generating distribution satisfies any of the assumptions of these models. For example, we use logistic regression models as working models for the propensity score and dropouts, but do not assume the conditional distribution has the functional form of a logistic regression model. We applied the weighted Generalized Linear Models (GLM) combining the three working models sequentially from the last visit to the baseline to derive the estimator of the treatment effect. The weights in GLM at each step are constructed using the predicted probabilities from the propensity score and dropout working models for the observed outcome values at that step. The propensity score working model is correctly specified by design in a randomized clinical trial, so that TMLE estimator is consistent if either the dropout or outcome working model is correctly specified. If all working models are correctly specified, then the TMLE estimator achieves the semiparametric efficiency bound. This makes TMLE a more robust estimator compared to MMRM or GLMM, especially under missing at random scenarios. A more detailed step-by-step procedure in constructing the TMLE estimator in longitudinal setting can be found in Rosenblum et al. (2018).

Non-parametric bootstrap (Efron and Tibshirani, 1994) was used to derive the variance for all the estimators.

## 4. Simulation Studies

### 4.1 *Simulation Design*

We compared the performance of MMRM, MMRM*, TMLE and GLMM by simulating randomized trials based on a real diabetes clinical trial data, which is a phase 3 randomized pivotal study with 3 post-baseline visits at week 4, 12, and 26, and the primary outcome of the study is the change in HbA1c from baseline to 26 weeks of treatment. Different dropout distributions under scenarios with zero or beneficial ($\Delta = -2\%$) treatment effect were considered in order to explore their impact on estimator bias, variance, mean squared error (MSE), relative mean squared error (RMSE) and coverage probability of the 95% BCa confidence interval based on 10,000 bootstrap samples.

Each simulated trial is generated first by resampling the patients (N=380) with replacement from the real data from those in both arms who completed the trial. By resampling patients from the trial, the relationship between the baseline variables and observed outcomes within the original study is retained. The baseline variables in this study include age, gender, body weight and HbA1c value at baseline.

Second, study arm assignment $A$ was reassigned for each simulated participant by an independent Bernoulli draw with probability 0.5 to receive active treatment ($A = 1$) vs. placebo ($A = 0$).

The third step only applies to simulated trials generated corresponding to the beneficial average treatment effect. With the continuous outcome, for each simulated participant in the treatment arm $A = 1$, her/his change in HbA1c at 4-, 12-, and 26-weeks were decreased by 1%, 1.5%, and 2%, respectively. With the binary outcome, for each simulated participant who was initially assigned A = 1 and Y = 0, we randomly replaced Y with 1 by an independent Bernoulli draw with probability 0.2, 0.25, and 0.3 at each visit respectively.

Lastly, patient dropout was generated under MCAR or MAR with a 15% missing at final visit (week 26). For the scenario of MCAR, patients have increasing dropout rates over time 5%, 10% and 15%. For the scenario of MAR, dropout depends on treatment and longitudinal history of HbA1c with the dropout rates of 5%, 10%, 15% in treatment arm, and 10%, 15%, 20% in placebo arm.

For each combination of the 2 average treatment effects and 2 dropout scenarios, 10,000 trials were simulated for both continuous and binary outcomes.

4.2 *Simulation Results*

A comparison of the unadjusted and adjusted estimators of the average treatment effect at final visit are presented in Table 1 for continuous outcome and in Table 2 for binary outcome. Table 1 shows consistent results across different dropout scenarios under both zero or beneficial treatment effect, where TMLE and MMRM* achieved greater improvement in bias and variance reduction over unadjusted estimator and MMRM. Among all adjusted estimators, MMRM* yielded the greatest performance with roughly a 122% gain in efficiency on average for estimating the average treatment effect comparing to the unadjusted estimator. TMLE achieved similarly high efficiency gain ranging from 112% to 123% when compared to unadjusted estimator. In this example, MMRM* that accounted for interactions between visits and baseline variables improved on MMRM with a relative efficiency gain ranging from 47% to 51%. Despite the fewer parameters estimated through MMRM, it is still likely to suffer a loss of efficiency by ignoring the time-varying correlation between the visits and the baseline variables.

[Table 1 about here.]

Table 2 displays the results for longitudinal binary outcome. TMLE yielded the lowest bias, variance and MSE in all cases, and it achieved an average efficiency gain of roughly 114% when compared to unadjusted estimator, and roughly 4-5% when compared to GLMM. Remember that the GLMM procedure involved multiple shots of model fitting with various covariance structures starting from the unstructured option until convergence. This inconvenience in practice further highlights the advantage of TMLE over GLMM in modeling longitudinal binary outcome.

[Table 2 about here.]

## 5. Clinical Trial Applications

Our data analyses for 28 clinical trial applications are summarized in Figure 1 and table S1 for longitudinal continuous outcome in HbA1c change from baseline, and in Figure 2 and table S2 for longitudinal binary outcome in whether a participant reaches the pre-specified HbA1c target (HbA1c < 7%). We used the variance ratio $\frac{V_{\text{unadj}}}{V_{\text{adj}}}$ to evaluate the performance of different adjusted estimators relative to the unadjusted estimator, where $V_{\text{unadj}}, V_{\text{adj}}$ indicate the variance of the unadjusted and adjusted estimators respectively. The higher the value is, the more precision it has in estimating the treatment constrasts. A value greater than 1.0 shows that this estimator is more precise than the unadjusted estimator. If assuming consistency of those estimators, the variance ratio is equivalent to the relative efficiency. Figure 1 and 2 show the variance ratio for all the adjusted estimators relative to the unadjusted estimator. Table S1 and S2 display the numeric results from data applications. Only variance and its ratio are shown in the tables for the purpose of confidentiality, but the estimates of the average treatment effect are very close for various estimators across datasets (data not shown here).

### 5.1 *Continuous Outcome*

Figure 1 showed that the adjusted estimators for average treatment effect are more precise than unadjusted estimator in most studies for continuous outcome, suggested by the fact that the majority of dots lie above the threshold of 1.0 for the variance ratio. Moreover, the MMRM* estimator (blue) seemed to achieve the highest efficiency gain in most cases (20 out of 28 studies). Compared to the MMRM estimator (green), the MMRM* estimator was at least as good in 26 out of 28 studies, and only slightly worse in 2 studies (study 15 and 24). Despite the recommendation of using MMRM* estimator (Mallinckrodt et al., 2008, 2020) in the analyses of longitudinal continuous outcome from randomized clinical trials, the MMRM estimator remains to be a commonly used choice. The simulation study and data applications in this paper provide compelling empirical evidence to support the use of MMRM* estimator which captures the varied correlation between the baseline variables and the outcomes over time. The MMRM* estimator is shown to provide more precision than the MMRM estimator. TMLE (orange) showed competitiveness to MMRM* estimator although on average it had a numerically less precision. Note that our performance evaluation focused on the estimated variance of the estimators only. The true data generating distribution remains unknown to us, therefore, we are not able to assess and compare the bias of different estimators. Although TMLE has comparable or larger variance than the MMRM* estimator, it may have smaller bias due to its doubly robust property under missing at random scenario. However, in the context of continuous-valued outcome, TMLE may not offer much advantage.

[Figure 1 about here.]

## 5.2 *Binary Outcome*

Figure 2 showed the variance ratio of GLMM (pink) or TMLE (orange) adjusted estimator over unadjusted estimator of risk difference for binary outcome at the last visit. Since GLMM could have different results due to the usage of various correlation structures, only the GLMM results with convergence and the highest efficiency gain in each dataset were displayed in Figure 2 and Table S2. Detailed GLMM results with regards to different correlation structures can be found in Table S3. In industry practice, GLMM usually starts with unstructured covariance matrix. If the result does not converge, a simpler covariance structure will be used, like AR1, compound symmetry until the convergence is achieved. This is a rather cumbersome procedure.

Figure 2 showed that the adjusted treatment effect estimators are better than unadjusted estimator in most studies (22 out of 28 studies) for binary outcome at the last visit. TMLE estimator is at least as good as GLMM estimator in 26 out of 28 studies, and only slightly worse in 2 cases. TMLE estimator avoids the cumbersome procedure from GLMM try-and-error approach in finding the appropriate covariance matrix that yields the convergent output. Note that we compared the TMLE with the best performing GLMM estimator, while in practice, only the first convergent GLMM result would be used where the sequence of covariance structures options will be pre-defined in the Statistical Analysis Plan (SAP); this may increase the benefit of using TMLE in estimating treatment contrasts with binary outcome. Again TMLE estimator comes with the doubly robust property that makes it a more robust estimator than GLMM and unadjusted estimator.

[Figure 2 about here.]

## 6. Practical Recommendations

We compared the commonly used MMRM/MMRM* and GLMM approaches versus the TMLE methods for estimating the average treatment effect using simulations and 28 randomized clinical trials from diabetes therapeutic area. The outcome of interest is either continuous-valued (change in HbA1c from baseline) or binary-valued (whether a participant reached a HbA1c target - HbA1c < 7%), collected longitudinally over the trial duration. Regardless of outcomes of interest, adjusted methods have been observed to achieve substantial precision gains by leveraging the available data when compared to unadjusted estimator. For modeling the longitudinal continuous outcome using mixed model approaches, precision in the estimated average treatment effect can be greatly improved by using MMRM* estimator that adds the interaction terms between visits and baseline variables. The MMRM* estimator has shown consistent results in both simulations and real data applications that it yielded higher precision gain compared to the MMRM estimator in the analyses of diabetes clinical trials. The MMRM* estimator also showed competitive or better precision gain than the TMLE estimator in many cases. Hence, we encourage the sponsors to use MMRM* estimator when estimating the average treatment effect with longitudinal continuous outcome in diabetes clinical trials, adding the visit-by-baseline interaction into the model as part of the fixed effects.

Concerning the binary outcome in a longitudinal setting, compared to the commonly used GLMM method, TMLE has shown greater advantages with reduced bias and variance in all simulated scenarios and being at least as good as GLMM estimator in 26 out of 28 diabetes clinical trial applications, yet without the cumbersome try-and-error covariance fitting procedure in GLMM. Moreover, the doubly robust property of TMLE estimator may offer more robustness to model misspecification in clinical trial applications. The GLMM results shown in simulations and data applications came from the best performing covariance structures with convergence after multiple model fitting, but in practice, only the first convergent GLMM result would be reported following a pre-defined sequence of covariance structure options. This further highlights the potentially greater benefit of using TMLE in estimating treatment effects with the binary outcome in a longitudinal setting.

Another advantage of considering TMLE approach, whether handling with longitudinal continuous-valued or binary-valued outcome, is that the sponsor can adjust for other related variables at each intermediate regression for the outcome modeling or dropout modeling. For example, in the setting of diabetes clinical trials, the participant dropout may be related to some safety events, like hypoglycemia. So the occurrence of the hypoglycemia event could be added to the dropout working model to better mimic this hypothesis. These post-baseline adjustment may further reduce the bias and improve the precision of the estimator.

Ma et al. (2022) compared the GLMM and the Multiple Imputation (MI) based approaches in the analyses of longitudinal binary outcome dichotomized from an underlying continuous variable, where MI-based approaches utilizing the information on the underlying continuous variables have been shown to yield more precise estimators than the GLMM. It remains to be an interesting topic to evaluate the performance of TMLE against such MI-based approaches as a future work.

We reported these findings from a systematic analysis of diabetes randomized clinical trials and we encourage the readers and/or sponsors to try out the approaches in different therapeutic areas.

**ORCID**

Lingjing Jiang https://orcid.org/0000-0001-8706-2850

Michael Rosenblum https://orcid.org/0000-0001-7411-4172

Yu Du https://orcid.org/0000-0003-0255-6283

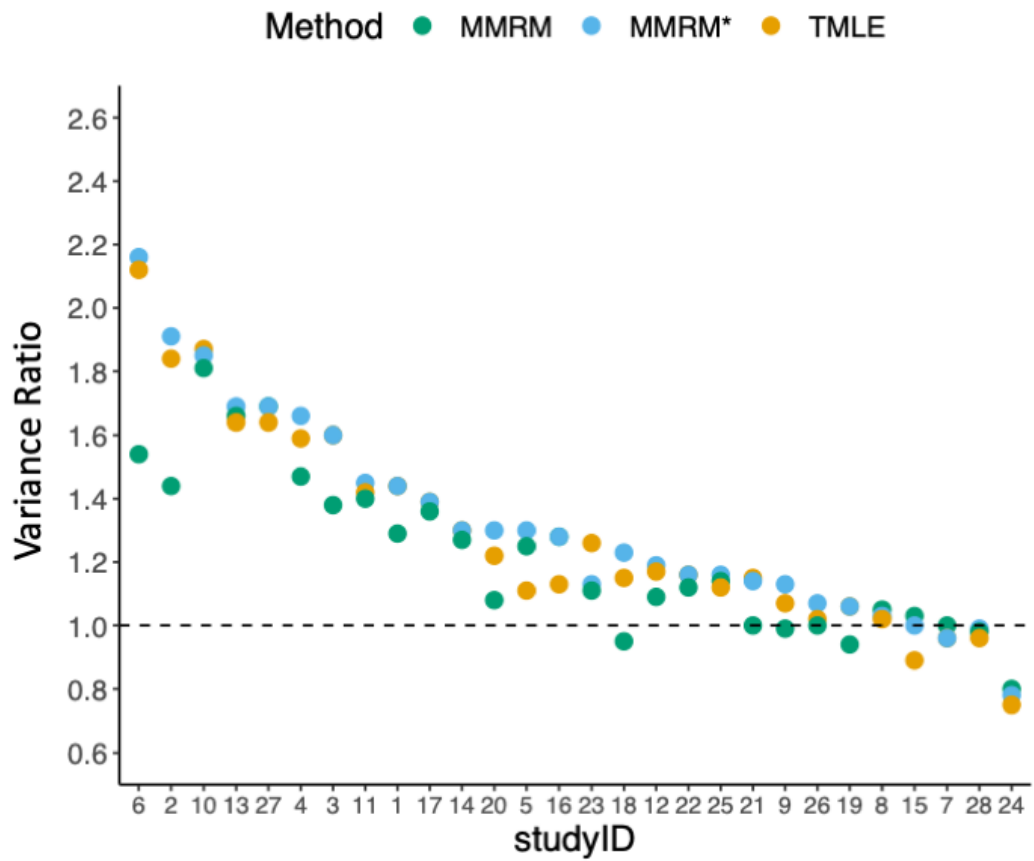

**Figure 1.** Method comparisons based on variance ratio to unadjusted estimator in clinical data set analyses for continuous outcome. The x-axis indicates the study ID for each dataset, and the y-axis shows the value of variance ratio. The dashed line suggests the value of variance ratio at 1.0, under which adjusted estimators have no advantage over unadjusted estimator. Different colors represent the results from different methods: MMRM (green), MMRM* (blue) and TMLE (orange).

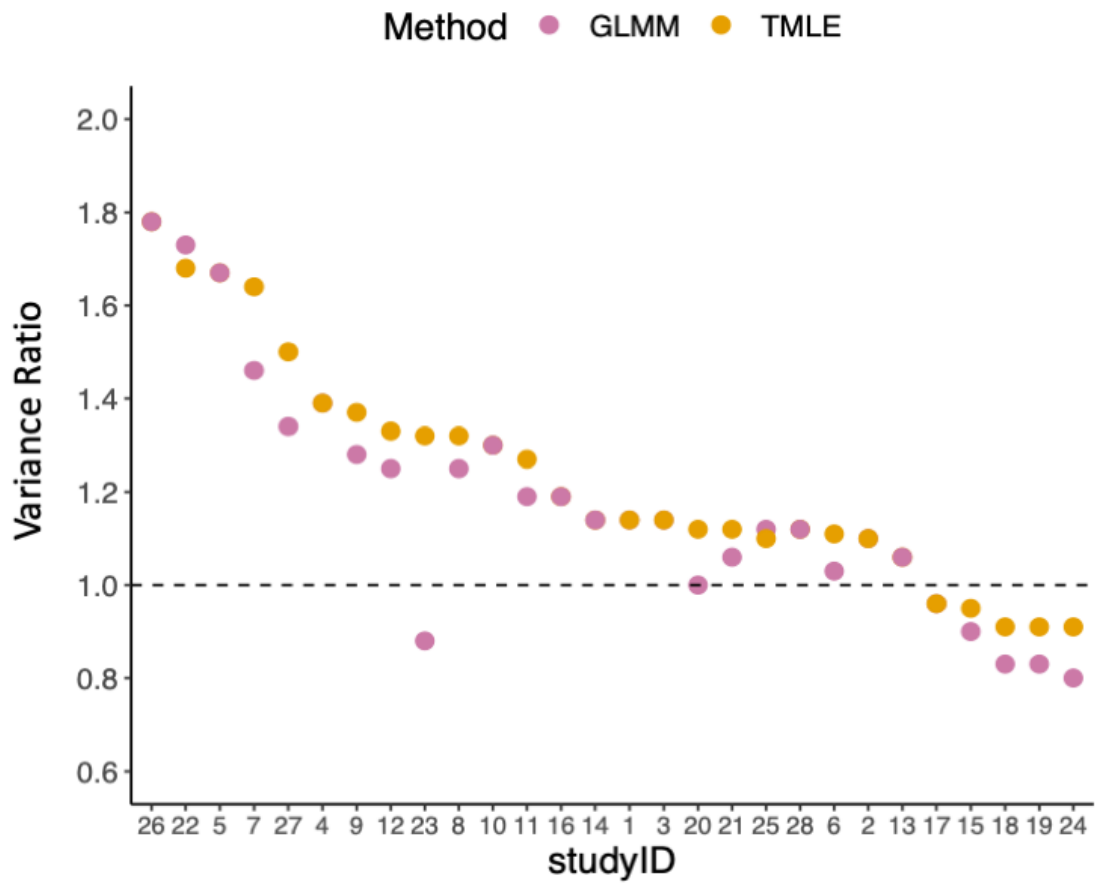

**Figure 2.** Method comparisons based on variance ratio to unadjusted estimator in clinical data set analyses for binary outcome. The x-axis indicates the study ID for each dataset, and the y-axis shows the value of variance ratio. The dashed line suggests the value of variance ratio at 1.0, under which adjusted estimators have no advantage over unadjusted estimator. Different colors represent the results from different methods: GLMM (pink), and TMLE (orange).

**Table 1**
*Simulation results for longitudinal continuous outcome*

| **Continuous outcome** | | | | | |
|---|---|---|---|---|---|
| Zero treatment effect ($\Delta = 0$) | | | | | |
| Dropout Scenario | Estimator | Bias | VAR | MSE | RMSE | CP |
| MCAR | Unadj | 0.0023 | 0.0126 | 0.0126 | 1.00 | 0.96 |
| | MMRM | 0.0021 | 0.0085 | 0.0085 | 1.48 | 0.96 |
| | TMLE | 0.0009 | 0.0059 | 0.0059 | 2.14 | 0.94 |
| | MMRM* | 0.0009 | 0.0058 | 0.0058 | 2.17 | 0.95 |
| MAR | Unadj | -0.0003 | 0.0127 | 0.0127 | 1.00 | 0.95 |
| | MMRM | -0.0029 | 0.0085 | 0.0086 | 1.48 | 0.95 |
| | TMLE | 0.0015 | 0.0060 | 0.0060 | 2.12 | 0.94 |
| | MMRM* | 0.0014 | 0.0058 | 0.0058 | 2.19 | 0.96 |
| Beneficial treatment effect ($\Delta = -2$) | | | | | |
| Dropout Scenario | Estimator | Bias | VAR | MSE | RMSE | CP |
| MCAR | Unadj | 0.0026 | 0.0130 | 0.0130 | 1.00 | 0.96 |
| | MMRM | 0.0024 | 0.0087 | 0.0087 | 1.49 | 0.96 |
| | TMLE | 0.0014 | 0.0060 | 0.0060 | 2.17 | 0.94 |
| | MMRM* | 0.0014 | 0.0059 | 0.0059 | 2.20 | 0.95 |
| MAR | Unadj | 0.0108 | 0.0135 | 0.0136 | 1.00 | 0.94 |
| | MMRM | -0.0041 | 0.0089 | 0.0089 | 1.53 | 0.95 |
| | TMLE | 0.0023 | 0.0061 | 0.0061 | 2.23 | 0.94 |
| | MMRM* | 0.0023 | 0.0059 | 0.0059 | 2.31 | 0.96 |

**Table 2**
*Simulation results for longitudinal binary outcome*

| **Binary outcome** | | | | | |
|---|---|---|---|---|---|
| Zero treatment effect ($\Delta = 0$) | | | | | |
| Dropout Scenario | Estimator | Bias | VAR | MSE | RMSE | CP |
| MCAR | Unadj | -0.0002 | 0.0062 | 0.0062 | 1.00 | 0.94 |
| | GLMM | -0.0004 | 0.0030 | 0.0030 | 2.07 | 0.94 |
| | TMLE | -0.0001 | 0.0028 | 0.0028 | 2.21 | 0.94 |
| MAR | Unadj | -0.0037 | 0.0062 | 0.0062 | 1.00 | 0.94 |
| | GLMM | -0.0006 | 0.0030 | 0.0030 | 2.07 | 0.94 |
| | TMLE | -0.0001 | 0.0029 | 0.0029 | 2.14 | 0.94 |
| Beneficial treatment effect ($\Delta = 0.3$) | | | | | |
| Dropout Scenario | Estimator | Bias | VAR | MSE | RMSE | CP |
| MCAR | Unadj | -0.0022 | 0.0059 | 0.0059 | 1.00 | 0.95 |
| | GLMM | -0.0062 | 0.0028 | 0.0029 | 2.03 | 0.95 |
| | TMLE | -0.0021 | 0.0028 | 0.0028 | 2.11 | 0.95 |
| MAR | Unadj | -0.0077 | 0.0058 | 0.0059 | 1.00 | 0.94 |
| | GLMM | -0.0087 | 0.0028 | 0.0029 | 2.03 | 0.94 |
| | TMLE | -0.0034 | 0.0028 | 0.0028 | 2.11 | 0.94 |